# Intrinsic Room Temperature Ferromagnetism in Boron-doped ZnO

X. G. Xu, H. L. Yang, Y. Wu, D. L. Zhang, S. Z. Wu, J. Miao, Y. Jiang[*]

State Key Laboratory for Advanced Metals and Materials, School of Materials Science and Engineering, University of Science and Technology Beijing, Beijing 100083, China

**Abstract**

We report room temperature ferromagnetism in boron-doped ZnO both experimentally and theoretically. The single phase $Zn_{1-x}B_xO$ films deposited under high oxygen pressure by pulsed-laser deposition show ferromagnetic behavior at room temperature. The saturation magnetization increases monotonously from 0 to 1.5 emu/cm$^3$ with the increasing of B component x from 0 to 6.8%. The first-principles calculations based on density functional theory demonstrate that the ferromagnetism in B-doped ZnO originates from the induced magnetic moments of oxygen atoms in the nearest neighbor sites to the B-Zn vacancy pair. The calculated total magnetic moment increasing tendency with B component is well consistent with experiments.

[*] To whom correspondence should be addressed. Email: yjiang@ustb.edu.cn

Dilute magnetic semiconductors (DMS) are promising candidates for spintronics devices due to their peculiar magnetic and semiconductor properties. In recent years, researchers have made great efforts to design and fabricate DMS, especially those with Curie temperature ($T_C$) above room temperature. Among all DMS candidates, ZnO has attracted wide interest, since it is predicted to be an ideal room-temperature DMS by Dietl et al. [1]. The magnetic moments of oxygen atoms around Zn vacancy are confirmed to be an origin of the ferromagnetic (FM) property in pure ZnO system [2,3]. However, the density of Zn vacancy is sensitive to experimental conditions and difficult to be controlled. Transition metal (TM) doping is a traditional method to obtain room temperature ferromagnetism in ZnO system [4,5,6,7,8,9,10,11,12]. However, the inconsistent experimental results raise a new problem of explaining the mechanism of ferromagnetism in TM-doped ZnO. Moreover, it is difficult to exclude the possibility that the ferromagnetism might be brought by a secondary phase of TM oxides or TM dopant clusters [13,14,15]. Consequently, many researchers turned their attention to non-TM doped ZnO and tried to provide undisputed intrinsic DMS. Pan et al. [16] theoretically predicted and experimentally realized room temperature ferromagnetism in carbon-doped ZnO films. A further work by Peng et al. [17] demonstrated the hole-induced mechanism in p-group element-doped FM ZnO system, which opens a new way for studying non-TM doped ZnO DMS.

In this letter, we report our experimental study on 2p-group element B-doped ZnO system. First-principles calculations based on the density functional theory justify the intrinsic ferromagnetism in the system is induced by oxygen atoms in the nearest neighbor sites to the B-Zn vacancy pair, which is quite different from the origin of the ferromagnetism in carbon-doped ZnO.

All the B-doped ZnO thin films were grown by pulsed-laser deposition (PLD) using a KrF

excimer laser operating at 300 mJ/pluse and 10 Hz. The targets were prepared by sintering mixed ZnO (99.99%) and $B_2O_3$ (99.99%) powders in air at 900 $^o$C for 30 min. The boron concentrations in the targets are 0%, 5% and 10%, respectively. The base pressure of PLD chamber was better than $5.0\times10^{-6}$ Pa. A series of $Zn_{1-x}B_xO$ thin films were deposited on quartz substrates at 400 $^o$C or 500 $^o$C under partial oxygen pressure of 10 Pa in order to ensure epitaxial growth and avoid oxygen vacancy during deposition. The boron component x in $Zn_{1-x}B_xO$ films was measured by electron probe microanalysis (EMPA). The thin films were characterized by X-ray diffraction (XRD) and X-ray photoelectron spectroscopy (XPS) to study their crystal structures and chemical valence states, respectively. Magnetic measurements were carried out using Alternating Gradient Magnetometer (AGM) at room temperature.

Boron is a light element which might be partially lost during the film deposition. So we used the EMPA technique to accurately determine the B component in $Zn_{1-x}B_xO$ thin films. It is demonstrated that the B atoms are uniformly distributed in B-doped ZnO films. The B component x is 4.3 and 4.2 at. % in the films deposited using the $ZnO/B_{0.05}$ target at 400 $^o$C and 500 $^o$C, respectively. And x is 6.8 and 6.0 at. % in those deposited using the $ZnO/B_{0.1}$ target at 400 $^o$C and 500 $^o$C, respectively. The B loss increases with both the increasing of B content in the target and the deposition temperature, indicating that only low B concentration is stable in $Zn_{1-x}B_xO$ system.

XPS was employed to decide the chemical valence states. For the $Zn_{1-x}B_xO$ (x=6.8%) film, as shown in Figure 1, all the indexed peaks are ascribed to C, O, Si and Zn elements, and no other TM ions can be detected. In the inset of Fig. 1, there is a weak peak at 192.7 eV derived from the B-O bonds, which indicates the doping B atoms are in the trivalent state. The XRD spectra of $Zn_{1-x}B_xO$ thin films are presented in Figure 2. It can be seen that there are only (002) and (004) peaks in all spectra, which indicates that both pure and B-doped ZnO films are single phase and have good epitaxy with the quartz substrates. It is noteworthy that the (002) peak of B-doped ZnO shows a right shift of ~0.5 deg relative to that of the pure ZnO film, corresponding to a lattice shrinking in *c* axis. There are three possible sites, octahedral interstice, Zn and O sites for B to reside in ZnO crystal. If the B occupies in the octahedral interstice, the lattice will be expanded. When $B^{3+}$ cations substitute the $O^{2-}$ ions, the Coulomb force between $Zn^{2+}$ and $B^{3+}$ cations will also expand the lattice. Moreover, because all the films were deposited under the oxygen pressure of 10 Pa, it is impossible to form oxygen-poor state. Combining the results of XRD and XPS, we can conclude that the B dopants tend to reside in Zn site. However, with the B component x increasing, there might be a small number of B atoms occupying the octahedral interstice to counteract the lattice distortion or form $B_2O_3$ phase at last.

The room temperature M-H loops of $Zn_{1-x}B_xO$ samples are shown in the inset of Figure 3. The curve of the x = 4.2% sample is omitted for clarity, since it is similar with that of the 4.3% one. It is clear from the inset that all the B-doped ZnO films take on FM property, which is enhanced with the increasing B concentration x. Whereas, the pure ZnO film is nonmagnetic, demonstrating the as-deposited ZnO are perfect films without Zn vacancy or TM-metal impurity. The normalized saturation magnetization ($M_s$) as a function of B component x is presented in Fig. 3. The $M_s$ increases monotonously with the increasing x. Considering the system purity demonstrated by XPS and XRD and the magnetic increasing trend with B content, it is reasonable to judge that the FM property in B-doped ZnO is intrinsic and determined by B concentration.

First-principles calculations based on the density functional theory were performed using Vienna ab initio Simulation Package (VASP) [18] to investigate the FM origination in the $Zn_{1-x}B_xO$ system. Exchange correlation interactions were described by the Perdew-Wang 91 (PW91) generalized gradient approximation (GGA) [19]. The projector-augmented wave (PAW) method [20] was used for the treatment of core electrons. For all systems, the kinetic cutoff energy was set as 400 eV. The bulk ZnO was calculated first using a 9 × 9 × 6 $k$-mesh. The optimized lattice parameters $a$ and $c$ are respectively 3.283 Å and 5.283 Å, in good agreement with experiment [21]. The models of B-doped ZnO were constructed by a 3 × 3 × 2 supercell of ZnO. One or two B atoms in the supercell correspond to 2.8% or 5.6% B concentration. We have calculated supercells with the B atoms occupying the Zn, O and interstice sites, respectively. Considering the trivalent B cations might induce $Zn^{2+}$ vacancy (denoted by $V_{Zn}$) in an oxygen rich environment, we have calculated Zn vacancy supercells of $Zn_{35}V_{Zn}O_{36}$ (ZnO+$V_{Zn}$), $Zn_{34}V_{Zn}BO_{36}$ (B+$V_{Zn}$), $Zn_{33}V_{Zn}B_2O_{36}$ (2B+$V_{Zn}$) and $Zn_{32}V_{Zn2}B_2O_{36}$ (2B+2$V_{Zn}$). In B-doped models, the $V_{Zn}$ is paired with B atoms. The 2B+2$V_{Zn}$ models with different distance between B-$V_{Zn}$ pairs are also calculated to study the B segregation. In all calculations, we have employed a 3 × 3 × 2 mesh of special $k$-points to perform the summation over the Brillouin Zone.

The most possible site for B should be tetrahedral or octahedral interstice in ZnO, due to the small radius of B. According to our calculation, the tetrahedral interstice is too small for B atom to be stable. When B occupies the octahedral interstice, the lattice parameters expand to 3.311 Å and 5.311 Å, and the model converges to nonmagnetic configuration, on the contrary of the experimental results. In the supercells of B occupying O sites, the lattice parameters are expanded as expected, and the B atoms tends to uniformly disperse in the supercell to lower the free energy. Each doping B atom has a magnetic moment of 2.9 $\mu_B$, however the antiferromagnetic (AFM) configuration is the ground state and is prior to FM at an energy of 0.16 eV. When the B atoms substitute Zn site, the supercell converges to a nonmagnetic ground state, even though the lattice shrinking to 3.245 Å and 5.230 Å is consistent with the experimental results. According to the analyses above, no vacancy-free B-doped ZnO supercell can converge to a stable FM ground state and have an expanded lattice simultaneously. So the FM mechanism of B-doped ZnO is quite different from that of C-doped system [16].

Since the films are all deposited under high oxygen pressure, the oxygen vacancy can hardly exist in the B-doped ZnO, whereas, the $Zn^{2+}$ vacancy might be induced by the trivalent $B^{3+}$ dopants. Models with $V_{Zn}$ are calculated, and their optimized lattice parameters, free energies and magnetic moments are listed in Table I. The ZnO+$V_{Zn}$ supercell shows lattice shrinking and FM ground state with a magnetic moment of 1.27 $\mu_B$, which is mainly attributed by the induced magnetic moment about 0.13 $\mu_B$ of the nearest neighbor oxygen atoms to the $V_{Zn}$ site and is consistent with the results of Kim et al. [3]. The conservation of charge in B-doped ZnO is represented by the model of 2B+$V_{Zn}$. As expected, this model has no net spin polarization because there is no unpaired electron. However, the charge unbalanced models B+$V_{Zn}$ and 2B+2$V_{Zn}$ have magnetic moments of 0.82 $\mu_B$ and 1.74 $\mu_B$, respectively, accompanied by a lattice shrinking with B content increasing. The magnetic moment per B-$V_{Zn}$ pair is smaller than that in ZnO+$V_{Zn}$, due to one hole less than the latter. The FM property comes from the induced magnetic moments of oxygen ranging from 0.09 $\mu_B$ to 0.14 $\mu_B$, which is smaller than that in ZnO+$V_{Zn}$ and indicates more occupied O-2p states. The induced magnetic moment of Zn is smaller than 0.03 $\mu_B$ which can be omitted. The calculated magnetic moments per supercell are presented in Fig. 3 together with the experimental values of $M_s$. The theoretically predicted magnetic moment of the B-doped ZnO increases monotonously with the B content, consistent with the experimental results.

Comparing the free energies of two $2B+2V_{Zn}$ models, we found that the $B-V_{Zn}$ pairs tend to occupy neighbor sites to decrease the free energy and lattice distortion. As a result, the $B_2O_3$ phase will emerge when B concentration exceeds the solid solubility.

Figure 4 shows the partial density of states (PDOS) of the O and Zn atoms nearest to the $B-V_{Zn}$ pair in $B+V_{Zn}$ supercell. The Fermi energy is set to zero. The O-p states are almost full occupied and nonmagnetic in pure ZnO. When the $B-V_{Zn}$ pair is introduced into the supercell, the majority spin states are still full occupied, while the minority spin states are partially filled, resulting in a net magnetic moment. Similar changes are shown in the Zn-d states, however, the magnetic moment is too small to influence the magnetic property of the system. The Zn-s state has no contribution to the FM property according to Fig. 4 (b). Therefore, the ferromagnetism of B-doped ZnO mainly comes from the induced magnetic moment of the O nearest to the $B-V_{Zn}$ pair.

It is noticeable that the theoretical magnetic moments are much larger than the experimentally observed values, despite their changing tendency via B content is consistent. As we calculated, in the B-doped ZnO systems, only the $B-V_{Zn}$ pairs can induce magnetic moment on the nearest coordinated oxygen atoms. In our experiments, though the oxygen pressure ensure a perfect oxygen lattice, the density of Zn vacancy is difficult to be controlled. There must be a number of B atoms tend to form the charge conservation structure of $2B+V_{Zn}$. Some other possibilities consuming effective B atoms, such as octahedral interstice cannot be neglected as well. The calculations also indicate that high concentration of B is impossible because of the segregation of $B-V_{Zn}$ pairs, which will form the $B_2O_3$ phase. Therefore to control the density of $B-V_{Zn}$ pairs should be a key point to improve the FM property of B-doped ZnO.

In summary, we demonstrated the intrinsic room temperature ferromagnetism in the B-doped ZnO films both experimentally and theoretically. The saturation magnetization ($M_s$) increases monotonously with the B component. First-principles calculations indicate that the ferromagnetism in B-doped ZnO is due to the induced magnetic moment of oxygen atom in the nearest neighbor site to the B-$V_{Zn}$ pair. All these results suggest the B-doped ZnO to be a promising candidate for DMS materials.

We thank Prof. M. B. A. Jalil for helpful discussion. This work was partially supported by the NSFC (Grant Nos. 50701005, 50831002, 50971025), the National Basic Research Program of China (Grant No. 2007CB936202), the Key Grant Project of Chinese Ministry of Education (No. 309006) and the Ph.D. Programs Foundation (Grant No. 20070008024) of the Chinese Ministry of Education.

(Springer-Verlag, Berlin, 1982), Vol. **17**.

TABLE I. Calculated lattice parameters, free energies and magnetic moments of B-doped ZnO supercells. Far and neighbor represent the distance between two B-$V_{Zn}$ pairs.

|  |  | ZnO | ZnO+$V_{Zn}$ | 2B+$V_{Zn}$ | B+$V_{Zn}$ | 2B+2$V_{Zn}$ Far | 2B+2$V_{Zn}$ Neighbor |
|---|---|---|---|---|---|---|---|
| Lattice paremeter (Å) | a | 3.283 | 3.245 | 3.252 | 3.227 | 3.183 | 3.209 |
|  | b | 3.283 | 3.245 | 3.241 | 3.245 | 3.241 | 3.239 |
|  | c | 5.283 | 5.228 | 5.212 | 5.187 | 5.122 | 5.165 |
| Free energy (eV) |  | -18.076 | -319.149 | -334.855 | -328.044 | -329.709 | -331.102 |
| Magnetic moment ($\mu_B$) |  | 0 | 1.27 | 0 | 0.82 | 1.85 | 1.74 |

Figure Captions

FIG. 1. XPS spectrum of 6.8% B-doped ZnO. The inset corresponds to the magnified B 1s peak.

FIG. 2. (Color online) XRD patterns of pure ZnO deposited at 500 $^o$C and B-doped ZnO films with different B content.

FIG. 3. (Color online) Room temperature $M_s$ of B-doped ZnO films and the calculated magnetic moment per supercell as a function of B content. The inset shows the M-H loops for pure ZnO and B-doped ZnO films measured at room temperature.

FIG. 4. (Color online) Partial density of states of (a) O-p, (b) Zn-s and (c) Zn-d electrons in pure ZnO (dashed line) and 2.8% B-doped ZnO (solid line).

Figure 1.   (X. G. Xu et al.)

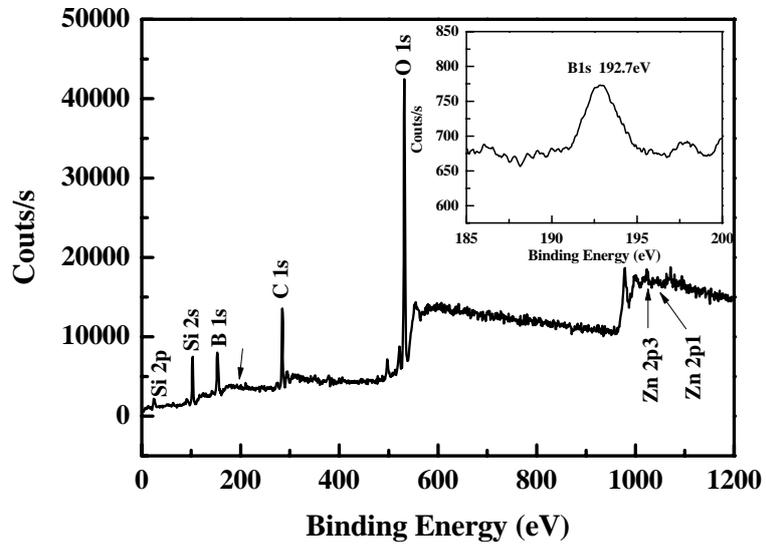

Figure 2. (X. G. Xu et al.)

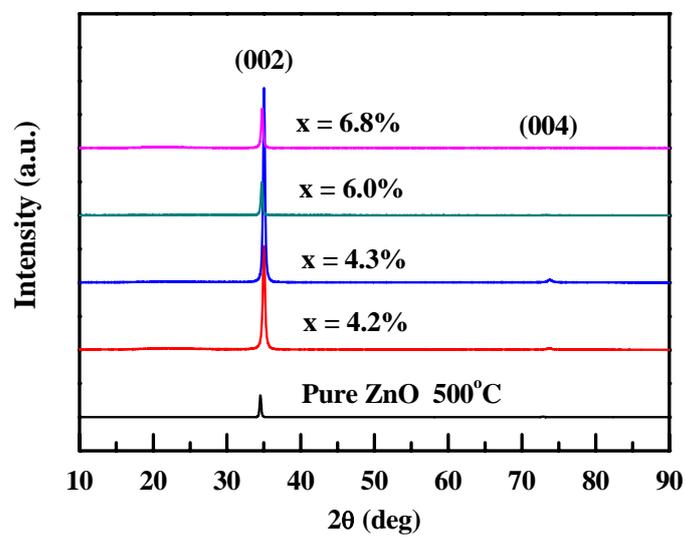

Figure 3. (X. G. Xu et al.)

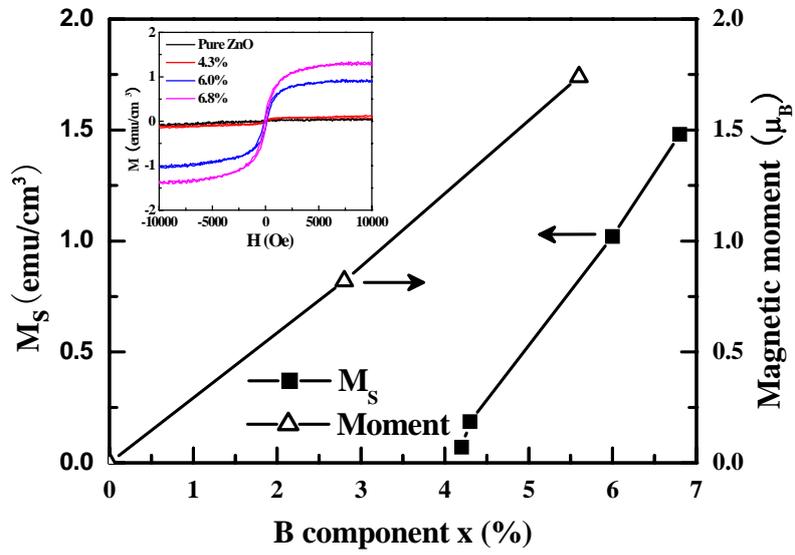

Figure 4. (X. G. Xu et al.)

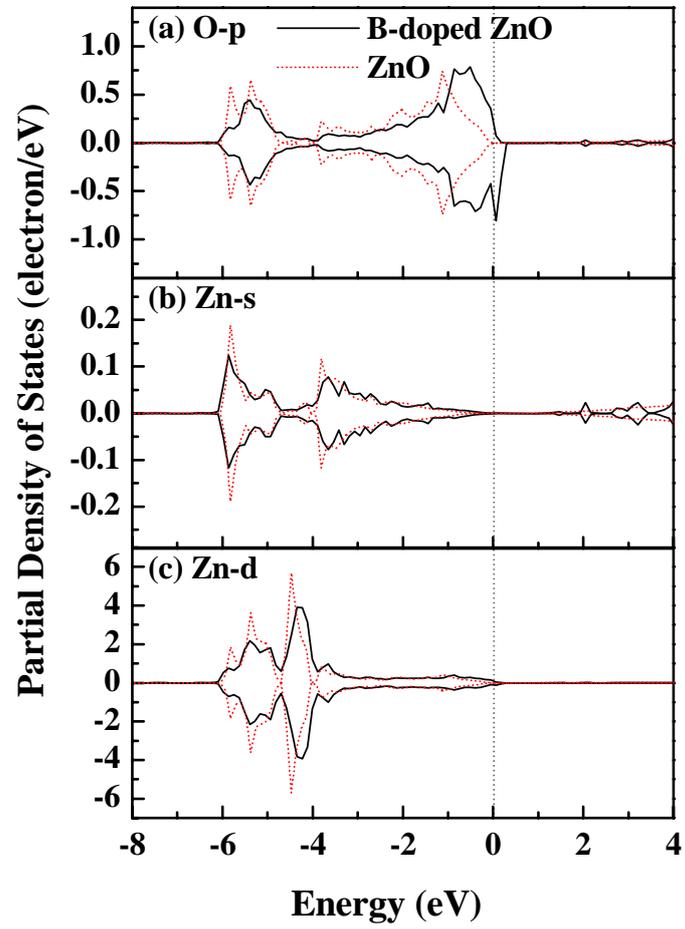